\newcommand{\bom}{\boldmath}
\documentstyle[12pt,iopconf,epsf]{article}
\begin{document}

\title{Neutrino Opacities at High Density and the Protoneutron Star Evolution}

\author{S Reddy$^\dag$\footnote{E-mail: reddy@nuclear.physics.sunysb.edu.},
J Pons$^\ddag$, M Prakash$^\dag$, and J M Lattimer$^\dag$}

\affil{$^\dag$Department of Physics \& Astronomy \\SUNY at Stony Brook, Stony
Brook, NY 11794-3800, USA}

\affil{$^\ddag$Departament d'Astronomia, Universitat de Val\'encia\\
    E-46100 Burjassot, Val\'encia, Spain}

\beginabstract
The early evolution of a protoneutron star depends both on the equation
of state and neutrino interactions at high density.  We identify the
important sources of neutrino opacity and through model calculations show
that in-medium effects play an important role in determining the neutrino mean
free paths. The effects due to Pauli-blocking and many-body correlations due
to strong interactions  reduce the neutrino cross sections by large factors
compared to the case in which these effects are ignored.   We discuss these
results in the context of neutrino transport in a protoneutron star.
\endabstract

\section{Introduction}
To date, calculations of neutrino opacities in dense matter have received
relatively little attention~\cite{Saw,IP,HW,RPL} compared to other physical
inputs such as the equation of state (EOS). The neutrino cross sections and
the EOS are intimately related. This relationship is most transparent in the
long-wavelength or static limit, in which the response of a system to a weak
external probe is completely determined by the ground state thermodynamics
(EOS).  Thus, in this limit, neutrino opacities consisitent with the EOS can be
calculated~\cite{Saw}. However, when the energy and momentum transfered by the
neutrinos are large, full consistency is often difficult to acheive in
practice.  Despite this, many salient features associated with an underlying
dense matter model may be incorporated in the calculation of the neutrino
opacities. In \S2, we describe how this is accomplished for a non-relativistic
potential model.  The effects of nucleon-nucleon correlations on the neutrino
mean free paths are calculated using the random phase approximation (RPA) in
\S3, where we show that the magnitudes of these many-body effects on both the
scattering and absorption reactions are large.  The total scattering cross
section in a multi-component system including the effects of correlation due to
both strong and electromagnetic interactions are presented in \S4 using a
relativistic formalism.  The implications of these results for neutrino
transport in a protoneutron star are in \S5.

\section{Neutrino Cross-Sections}
In the non-relativistic limit for the baryons, the cross-section per unit
volume for the two body reaction $\nu + B_2 \rightarrow l + B_4$ is given
by~\cite{RPL}
\begin{eqnarray}
\frac{\sigma(E_1)}{V}= \frac{G_F^2}{4\pi^2}({\cal V}^2+3{\cal A}^2)
\int_{-\infty}^{E_1} dq_0~
\frac{E_3}{E_1} (1-f_3(E_3))
\int_{\left|q_0\right|}^{2E_1-q_0} dq~ q S(q_0,q) \,.
\label{nr}
\end{eqnarray}
The particle labels are: 1:=incoming neutrino, 2:=incoming baryon, 3:=outgoing
lepton (electron or neutrino), and 4:=outgoing baryon. The energy transfer to
the baryons is denoted by $q_0=E_1-E_3$ and the momentum transfer
$q=|\vec{k_1}-\vec{k_2}|$.  Eq.~(1) describes both the charged and neutral
current reactions with the appropriate substitution of vector and axial vector
coupling constants, ${\cal V}$ and ${\cal A}$,  respectively~\cite{RPL}.
The factor $(1-f_3)$ accounts for Pauli blocking of the final state lepton.
The response of the system is characterized by the function $S(q_0,q)$, often
called the dynamic form factor. In \S3, we consider the  modifications
required in the presence of spin and isospin  dependent forces.

At the mean field level, the function $S(q_0,q)$ may be evaluated exactly  if
the single particle dispersion relation is known~\cite{RPL}.   We illustrate
this using a potential model. In this case, if we retain only a quadratic
momentum dependence, the single particle spectrum closely resembles that of
a free gas and is given by
\begin{eqnarray}
E_i(p_i)=\frac{p_i^2}{2M_i^*}+U_i \,, \quad i=n,p \,.
\label{nrspec}
\end{eqnarray}
\begin{figure}[t]
\vspace*{-0.1in}
\begin{center}
\epsfxsize=6.2in
\epsfysize=4.in
\epsffile{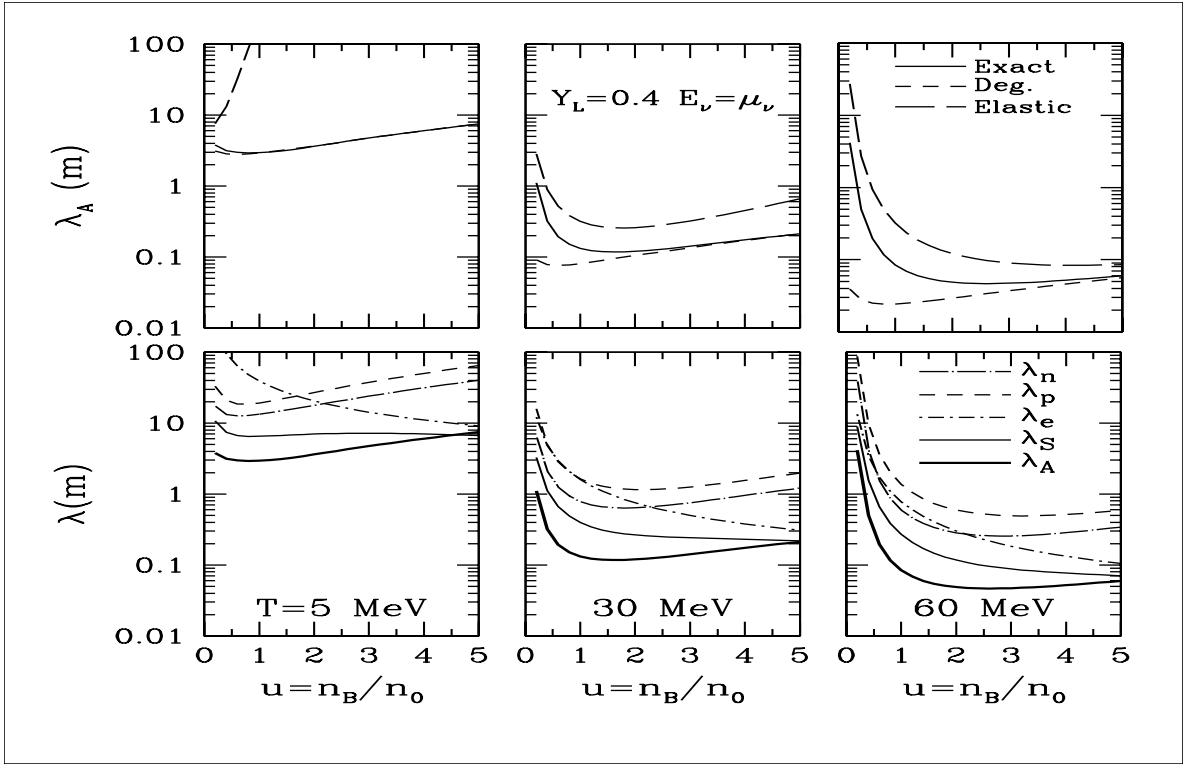}
\caption[]{\footnotesize Absorption (top panels) and scattering (bottom
panels) mean free paths in $\beta$-equilibrated stellar matter.}
\end{center}
\vspace*{-0.4in}
\end{figure}
The single particle potentials $U_i$ and the effective masses $M_i^*$
are density dependent.  Because the functional dependence of the spectra on the
momenta is similar to that of the noninteracting case, it is possible to obtain
an  analytic expression for the dynamic form factor $S(q_0,q)$~\cite{RPL}.
Explicitly,
\begin{eqnarray}
S(q_0,q)&=& \frac{M_2^*M_4^*T}{\pi q}~\frac{\xi_--\xi_+}{1-\exp(-z)} \,
\label{nricstruc}
\end{eqnarray}
where
\begin{eqnarray*}
\xi_\pm & = & \ln\left[\frac{1+\exp((e_\pm-\mu_2+U_2)/T)}
{1+\exp((e_\pm+q_0-\mu_4+U_2)/T)}\right] \nonumber \\
e_{\pm} & = & \frac{2q^2}{2M^*\chi^2}\left
[\left(1+\frac{\chi M_4^*c}{q^2}\right) \pm
\sqrt{1+\frac{2\chi M_4^*c}{q^2}}~\right] \,,
\end{eqnarray*}
with $\chi = 1- (M_4^*/M_2^*)$ and $c=q_0+U_2-U_4-(q^2/2M_4^*)$. The factor
$U_2-U_4$ is the potential energy gained in converting a particle of species
``2'' to a particle of species ``4''. In neutral current reactions, the initial
and final state particles are the same; hence, the strong interaction
corrections are due only to $M_2^*$. For the charged current reactions,
modifications due to interactions are twofold.  First, the difference in the
neutron and proton single particle potentials appears in the response function
and also in $\hat\mu=\mu_n-\mu_p$.  Second, the response depends upon the
nucleon effective masses.

Results obtained using Eq.~(3) are shown in Fig.~1, where electron neutrino
absorption (top panels) and scattering (bottom panels) mean free paths are
shown in  matter containing nucleons and leptons. Various limiting forms  are
also shown for comparison. The charged current mean free paths are shown as
thick solid lines in the bottom panel.  These results demonstrate that effects
due to kinematics, Pauli blocking, mass, and energy shifts are quantitatively
important~\cite{RPL}.

\section{Effects of Correlations}
The random phase approximation (RPA) is particularly suited to investigate the
role of particle-hole interactions on the collective response of matter. In
this approximation, ring diagrams are summed to all orders.  In terms of this
more general response, the differential cross-sections are given by
\begin{eqnarray}
\frac{1}{V}\frac{d^3\sigma}{d^2\Omega~dE_3}=\frac{G_F^2}{\pi}
(1-f_3(E_3))~~{\sc R}(q_0,q) \,,
\end{eqnarray}
where ${\sc R}(q_0,q)$ describes the system's response.  For the
neutral current reactions~\cite{IP}
\begin{eqnarray}
{\sc R}_S(q_0,q)=\left[c_V^2~(1+\cos{\theta})S_{00}{(q_0,q)}
+c_A^2~(3-\cos{\theta})S_{10}{(q_0,q)}\right] \,.
\end{eqnarray}
$S_{00}$ and $S_{10}$ are the density-density and spin-density response
functions. Similarly, the charged current response  may be written
in terms of the isospin-density  and the spin-isospin density response
functions as
\begin{eqnarray}
{\sc R}_A(q_0,q)=\left[g_V^2~(1+\cos{\theta})S_{01}{(q_0,q)}
+g_A^2~(3-\cos{\theta})S_{11}{(q_0,q)}\right] \,.
\end{eqnarray}
\subsection{Neutral Currents}
\begin{figure}[t]
\vspace{-0.1in}
\begin{center}
\epsfxsize=5.5in
\epsfysize=3.5in
\epsffile{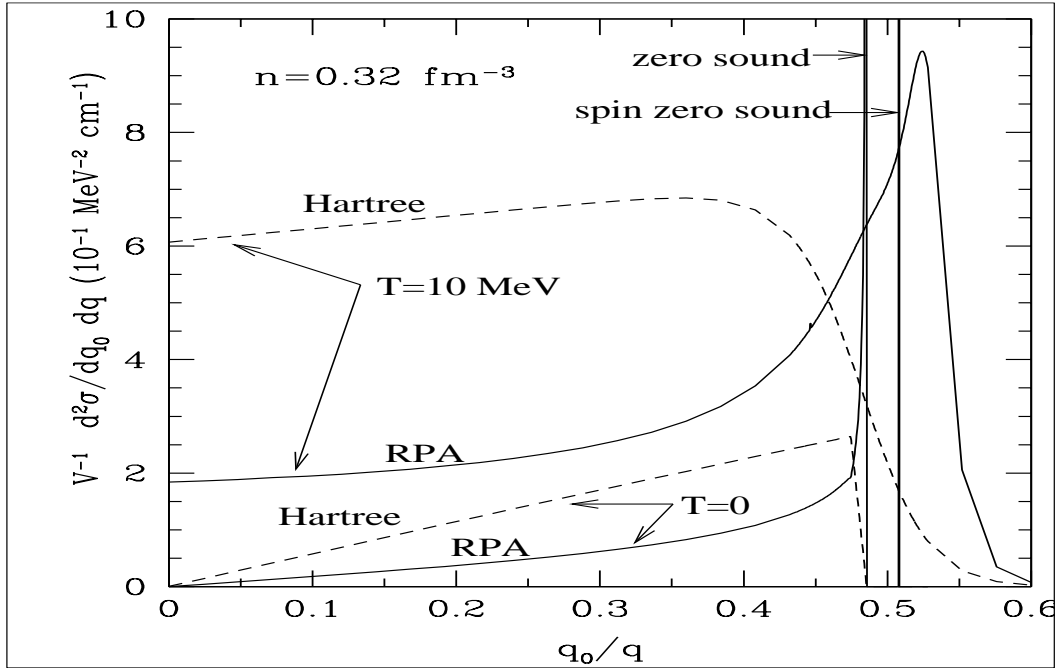}
\caption[]{\footnotesize Neutrino scattering cross sections versus
$q_0/q$ in pure neutron matter for $q$=10 MeV and a neutrino energy of
$E_{\nu}= 50$ MeV.}
\vspace*{-0.4in}
\end{center}
\end{figure}
Neutrino scattering on neutrons is the dominant source of scattering opacity.
We therefore begin by considering the response in pure neutron matter.  For a
one component system, the RPA response functions are given by
\begin{eqnarray}
S_{ij}(q_0,q)&=&\left[ \frac{1}{1-\exp(q_0/T)}\right]
~\frac{Im~\Pi^{0}(q_0,q)}{\epsilon_{ij}} \nonumber \\
\epsilon_{ij}&=&[1-V_{ij} Re~\Pi^{0}(q_0,q)]^2 +
[V_{ij} Im~\Pi^{0}(q_0,q)]^2 \,.
\end{eqnarray}
The dielectric screening function $\epsilon_{ij}$ is the modification
introduced by the RPA to Eq.~(1).  The zeros of $\epsilon_{ij}$
correspond to collective excitations, such as zero-sound and spin zero-sound.
The potential $V_{ij}$, which measures the strength of the particle-hole
interaction in the medium, is a function of density, temperature, $q_0$, and
$q$. Both the energy and momentum transferred in the particle-hole channel are
of order $T$. Since the particle-hole interaction is short ranged
($\sim$1/meson mass), explicitly density dependent interactions play a
major role in determining the magnitude of $V_{ij}$. Momentum dependent
interactions introduce additional structure to $V_{ij}$, which in turn alters
the structure of $S_{ij}$. We will consider these latter  effects separately.

For situations in which $q/k_{F_i}<<1$ and $T/\mu_i<<1$, the quasi-particle
interaction may be obtained using Fermi-liquid theory.  Here, the
quasi-particle interaction is given by
$V_{ij}=(\delta^2E/\delta n_{i}\delta n_{j})$, and is usually
expressed in terms of the Fermi-liquid parameters. For pure neutron matter, the
force in the spin independent channel is given by $F_0$, while the force in the
spin dependent channel is given by $G_0$.  Thus, $V_{00}=F_0/N_0$ and
$V_{10}=G_0/N_0$, where $N_0$ is the density of states at the Fermi surface
\cite{Mig}.  We employ the results of B\"{a}ckmann \etal~\cite{Back} who have
calculated $F_0$ and $G_0$ in  pure neutron matter for the densities of
interest here.

In Fig.~2, the differential cross sections for $\nu+n\rightarrow \nu+n$ are
shown. The effects due to correlations lead to
significant reductions at small $q_0$, since the interaction is repulsive and
the excitation of collective modes enhances the response at large
$q_0$. The well defined zero-sound and spin zero-sound seen at $T=0$ are damped
at finite temperatures. The contribution from the low $q_0$ region dominates
the total cross-section due to final state blocking.  The presence of a
repulsive particle-hole force acts to reduce the neutrino cross sections
for neutrino energies of order $T$.

The composition of charge neutral, $\beta$-equilibrated stellar matter depends
on the EOS of dense matter, which contains an admixture of protons and
electrons. The correlations due to interactions between the different particle
species and  electromagnetic correlations also play an important
role.  In \S4,  we discuss a relativistic framework to
describe the RPA response of a mixture of neutrons, protons, and electrons.

\subsection{Charged Currents}
\begin{figure}[t]
\vspace{-0.1in}
\begin{center}
\epsfxsize=5.5in
\epsfysize=3.5in
\epsffile{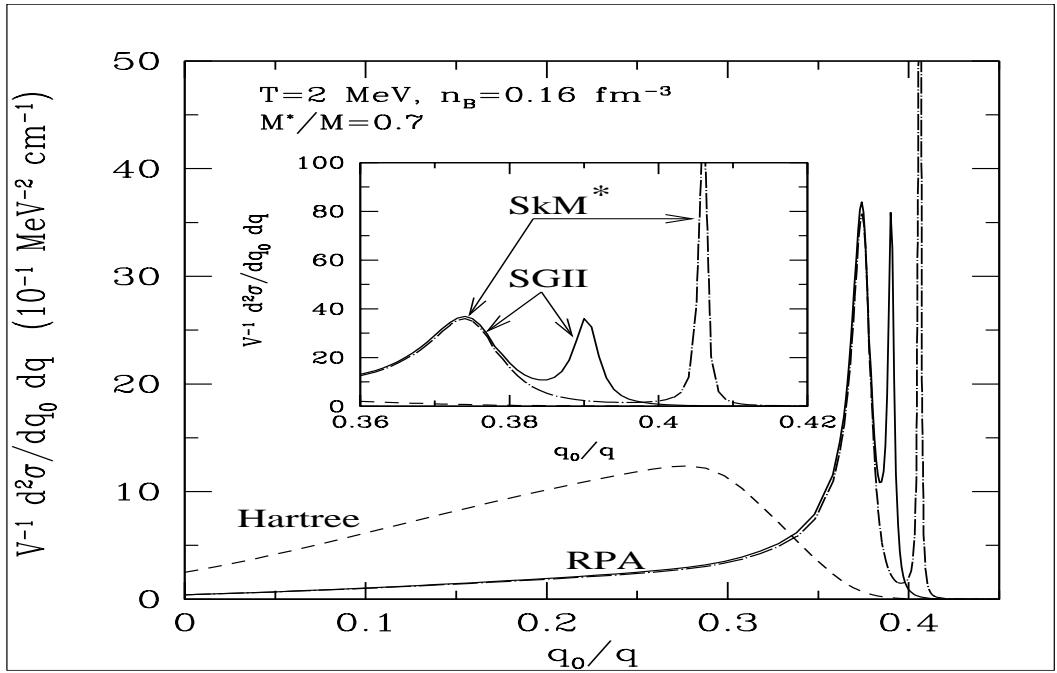}
\caption[]{\footnotesize Charged current reaction cross sections in symmetric
nuclear matter.}
\end{center}
\vspace*{-0.4in}
\end{figure}
The charged current reaction is 
kinematically different from the neutral current
reaction, since the energy and momentums transfers  are
not limited only by the matter temperature.  The energy transfer is typically
of order $\hat\mu=\mu_n-\mu_p$. The charged current probes smaller distances;
hence, the $q_0$ and $q$ dependencies of the particle-hole force are likely to
play an important role. In symmetric nuclear matter $\hat\mu=0$,
  and therefore, 
the situation is very similar to the neutral current case. Since the collective
response of laboratory nuclei is well established in nuclear physics, we
begin by considering the response of symmetric nuclear matter to a charged
current weak probe. Here the strength parameters $V_{01}$ for the spin
independent iso-spin channel and the $V_{11}$ for the spin dependent iso-spin
channel are required. In nuclear matter, the particle-hole force (retaining
only the $l=0$ terms) is given by \cite{Mig}
\begin{eqnarray}
{\sc F}(k_1,k_2)=N^{-1}_0 [F_0 + G_0 \bom{\sigma_1 \sigma_2}+
\bom{\tau_1 \tau_2} (F'_0+G'_0(\bom{\sigma_1 \sigma_2})] \,.
\end{eqnarray}
For the charged current reaction, isospin and charge are transferred along the
particle-hole channel; hence, only the last two terms in Eq.~(8) contribute. The
potentials required to calculate the RPA response in Eq.~(7) are given by
$V_{10}=2F'_0/N_0$ and $V_{11}=2G'_0/N_0$, where the factor two arises due to
isospin considerations. The Fermi-liquid parameters may be calculated from an
underlying dense matter model, as for example, a Skyrme model, which
successfully describes the low-lying excitations of 
nuclei~\cite{BT}. In Fig.~3, 
the response of symmetric nuclear matter to the charged current probe is shown.
Results are for two different parameterizations of the Skyrme force,
SGII and SkM$^*$.  The excitation of giant-dipole and Gamow-Teller resonances
shifts the strength  to large $q_0$, and as in the case of
scattering, the low $q_0$ response is significantly suppressed.

\begin{figure}[h]
\vspace{-0.1in}
\begin{center}
\epsfxsize=6.in
\epsfysize=3.5in
\epsffile{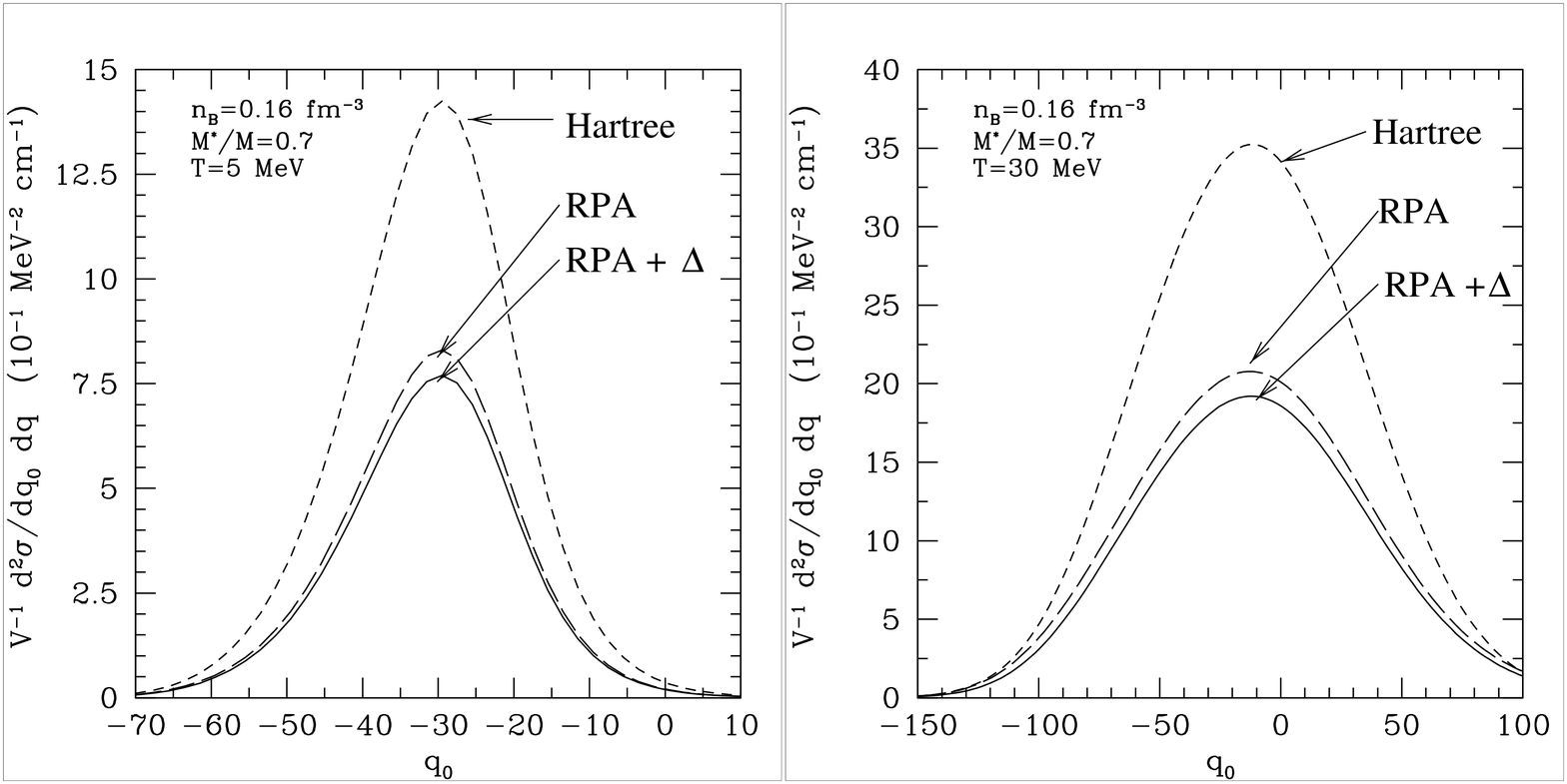}
\caption[]{Neutrino absorption cross sections in asymmetric matter.}
\end{center}
\vspace*{-0.4in}
\end{figure}
In asymmetric matter, the energy and momentum transfers are large.
Therefore, the momentum dependence of the particle-hole force becomes
important. At high momentum transfer, the conservation of the vector
current implies that the response function $S_{10}$ is not strongly
modified. Further, since its contribution to the differential cross
section is roughly  three times smaller than the spin-dependent
response function $S_{11}$, we focus on the Gamow-Teller part and
assume that the Fermi matrix elements are not screened.  The
iso-vector interaction in the longitudinal channel arises due to $\pi$
exchange, and in the transverse channel due to $\rho$ meson
exchange. In addition, to account for the large repulsion observed,
Migdal~\cite{Mig} introduced screening in this channel through the
parameter $g'$.  This form for the particle-hole interaction has been
successful in describing a variety of nuclear phenomena. The
longitudinal and transverse potentials in the $\pi + \rho + g'$ model
are given by \cite{Oset} 
\begin{eqnarray}
V_{L}(q_0,q) &=&
\frac{f^2_{\pi}}{m^2_{\pi}}
\left(\frac{\bf{q}^2}{q^2_0-\bf{q}^2-m^2_{\pi}}F^2_{\pi}(q) + g'\right)
\nonumber \\
V_{T}(q_0,q) &=&
\frac{f^2_{\pi}}{m^2_{\pi}}
\left(\frac{\bf{q}^2~C_{\rho}}{q^2_0-\bf{q}^2-m^2_{\rho}}F^2_{\rho}(q)
+ g'\right) \,,
\end{eqnarray}
where $F_{\pi}=(\Lambda^2-m^2_{\pi})/(\Lambda^2-q^2)$ and
$F_{\rho}=(\Lambda_{\rho}^2-m^2_{\rho})/(\Lambda_{\rho}^2-q^2)$ are the $\pi
NN$ and $\rho NN$ form factors.  Numerical values used are
$C_\rho=2,g'=0.7, \Lambda=1.4~{\rm GeV}$, and $\Lambda_\rho=2~{\rm GeV}$.
The RPA response function then takes the form
\begin{eqnarray}
S_{11}(q_0,q)&=&\left[ \frac{1}{1-\exp(q_0/T)}\right] ~Im~\Pi^{0}(q_0,q)
\left(\frac{1}{3 \epsilon_{L}} + \frac{2}{3 \epsilon_{T}}\right) \nonumber \\
\epsilon_{L,T} &=& [1-2 V_{L,T} Re~\Pi^{0}(q_0,q)]^2 +
[2 V_{L,T} Im~\Pi^{0}(q_0,q)]^2
\end{eqnarray}
The cross sections in charge neutral stellar matter with a fixed
lepton fraction $Y_L=0.4$ are shown in Fig.~4. The RPA screening reduces the
cross sections by about a factor of two. The inclusion of virtual $\Delta$-hole
excitations reduces the cross sections further. Unlike in the case of ordinary
$\beta$ decay, where the $\Delta$-hole contribution is important, the role
of these excitations is marginal here due to the large energy and momentum
transfers.

\section{Relativistic Treatment}
\begin{figure}[h]
\vspace{-0.1in}
\begin{center}
\epsfxsize=5.5in
\epsfysize=3.5in
\epsffile{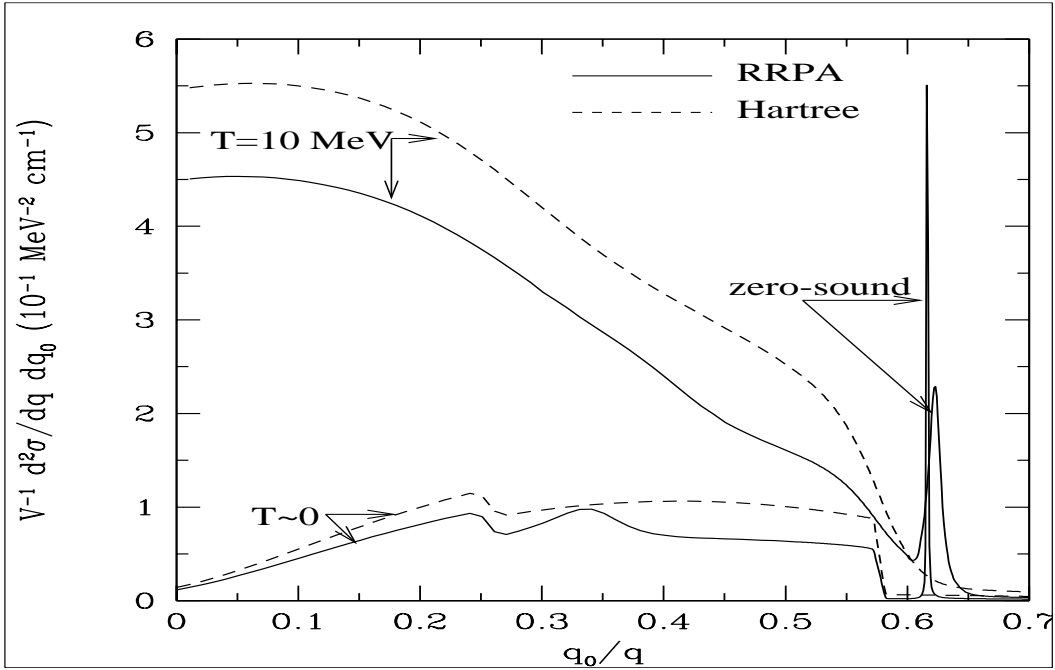}
\caption[]{Neutrino scattering cross sections in a relativistic approach.}
\end{center}
\vspace*{-0.4in}
\end{figure}
At high density the baryons become increasingly relativistic and thus a
relativistic description may be more appropriate.   Such a description also
allows us to treat the baryons and the electrons on an equal footing. The basic
formalism and some illustrative results may be found in \cite{HW,RPL}. Here we
present results for a system of neutrons, protons, and electrons at finite
temperature. The ground state is described by a field-theoretical model at the
mean field level, in which isoscalar $\sigma,\omega$ and
isovector $\rho$ meson exchanges modify the in-medium baryon propagators.  We
calculate the Hartree response by accounting for these modifications.
Correlations are incorporated through the relativistic random phase
approximation (RRPA), where the particle-hole interaction is mediated by the
$\sigma$ and $\omega$ mesons between the baryons, and electromagnetic
interactions between protons and electrons are mediated by photons. Fig. 5
shows the results for different temperatures at a baryon density of $0.32~{\rm
fm}^{-3}$. Compared to the non-relativistic models, the screening effects are
small, the dominant suppression arising  mainly due to density dependent
nucleon effective masses.

\section{Discussion}
We have highlighted the influence of correlations and collective
phenomena on the neutrino opacities in dense matter. Our findings here 
indicate that the neutrino cross sections are significantly reduced, and
the average energy transfer in neutrino-nucleon interactions is increased
due to the presence correlations in the medium.   Several
improvements are necessary before we can assess the influence of these
results on the macrophyscial evolution of a protoneutron star.  Among
the most important of these are calculations that provide (1) the
dynamic form factor, (2) the particle-hole and particle-particle
interactions, (3) the renormalization of the axial charge, and, (4)
the means to assess the role of multi-pair excitations, in charge
neutral, beta-equilibrated dense matter at finite temperature.  While
investigations along these directions are in progress (see also 
\cite{BS}), some general trends may be anticipated. The many-body effects 
studied here, including the improvements listed above, suggest considerable 
reductions in the opacities compared to the free gas estimates often employed 
in many applications. In the particular instance of the
early evolution of a protoneutron star, the suggested modifications
imply shorter time scales over which the deleptonization and cooling
occur.  Since mean free paths $\lambda$ are much less than the stellar
radius $R$, evolution is via diffusion.  To order of magnitude,
therefore, the timescale $\tau\propto1/\lambda$.  However, there are
important feedbacks between these evolutionary consequences and the
underlying EOS, in particular the specific heat of multicomponent
matter, which have to be studied before firm conclusions may be drawn.
\section*{Acknowledgments}
We thank G. E. Brown for useful discussions. This work was supported in part by
the U. S. Department of Energy under contract DOE/DE-FG02-88ER-40388, and the 
NASA grant NAG 52863.

\end{document}